\documentclass[runningheads]{llncs}
\usepackage{graphicx}
\usepackage{booktabs}
\usepackage[colorlinks=true,
            linkcolor=blue,
            filecolor=blue,      
            urlcolor=blue,       
            citecolor=blue
            ]{hyperref}
\usepackage{multicol,listings,multirow}
\usepackage[table]{xcolor}
\usepackage{caption}
\usepackage{lipsum}
\usepackage{tikz}
\usepackage{float}
\usepackage{diagbox}
\usepackage{amsmath}
\usepackage{amssymb}
\usepackage{wrapfig}
\usepackage{array}
\usepackage{booktabs}
\usepackage{array}
\usepackage{color}
\usepackage[marginal]{footmisc}
\usepackage[misc]{ifsym}
\usepackage{appendix}

\begin{document}
%

\title{Unifying and Personalizing Weakly-supervised Federated Medical Image Segmentation via Adaptive Representation and Aggregation}


%
\titlerunning{FedICRA for Weakly Federated Medical Image Segmentation}
%

\vspace{-0.8cm}
\author{Li Lin\inst{1,2,3}\and
Jiewei Wu\inst{1} \and
Yixiang Liu\inst{1} \and
Kenneth K. Y. Wong\inst{2} \and
Xiaoying Tang$^{1, 3(\textrm{\Letter)}}$}

\vspace{-0.2cm}
\institute{Department of Electrical and Electronic Engineering, Southern University of Science and Technology, Shenzhen, China\\ 
\email{tangxy@sustech.edu.cn}\and
Department of Electrical and Electronic Engineering, The University of Hong Kong, Hong Kong SAR, China
\and
Jiaxing Research Institute, Southern University of Science and Technology, \\
Jiaxing, China}
\vspace{-0.9cm}

\authorrunning{L. Lin et al.}
%

%
%
\maketitle              
\setlength{\skip\footins}{0.8cm}
\setlength{\footnotesep}{0.2cm}

\footnote{L. Lin, J. Wu and Y. Liu contributed equally to this work.}
\widowpenalty=5
\clubpenalty=5
\setlength\abovedisplayskip{0.21cm}
\setlength\belowdisplayskip{0.2cm}
\vspace{-0.8cm}
\begin{abstract}
Federated learning (FL) enables multiple sites to collaboratively train powerful deep models without compromising data privacy and security. The statistical heterogeneity (e.g., non-IID data and domain shifts) is a primary obstacle in FL, impairing the generalization performance of the global model. 
Weakly supervised segmentation, which uses sparsely-grained (i.e., point-, bounding box-, scribble-, block-wise) supervision, is increasingly being paid attention to due to its great potential of reducing annotation costs. However, there may exist label heterogeneity, i.e., different annotation forms across sites. 
In this paper, we propose a novel personalized FL framework for medical image segmentation, named FedICRA, which uniformly leverages heterogeneous weak supervision via adapt\textbf{I}ve \textbf{C}ontrastive \textbf{R}epresentation and \textbf{A}ggregation. Concretely, to facilitate personalized modeling and to avoid confusion, a channel selection based site contrastive representation module is employed to adaptively cluster intra-site embeddings and separate inter-site ones. To effectively integrate the common knowledge from the global model with the unique knowledge from each local model, an adaptive aggregation module is applied for updating and initializing local models at the element level. Additionally, a weakly supervised objective function that leverages a multiscale tree energy loss and a gated CRF loss is employed to generate more precise pseudo-labels and further boost the segmentation performance. 
Through extensive experiments on two distinct medical image segmentation tasks of different modalities, the proposed FedICRA demonstrates overwhelming performance over other state-of-the-art personalized FL methods. Its performance even approaches that of fully supervised training on centralized data. Our code and data are available at \url{https://github.com/llmir/FedICRA}.
\keywords{Personalized federated learning  \and Heterogeneous weak supervision \and Contrastive representation \and Adaptive aggregation.}
\end{abstract}
%
%
%

\section{Introduction}


Deep learning methods have been widely adopted in many computer vision and medical image analysis tasks due to their impressive performance \cite{lin2021bsda,cheng2021prior,lin2021blu}. However, as a data-driven approach, its performance is highly reliant on the quantity of accessible data. Data sharing across institutions, particularly medical sites, is often infeasible due to regulatory constraints and privacy concerns regarding user/patient data \cite{goddard2017eu}. In such cases, federated learning (FL) \cite{konevcny2016federated} has come to the fore and is receiving growing interest from the community as it allows different centers to collaboratively train a powerful global model without the need for sharing or centralizing data. As an illustration, FedAvg \cite{mcmahan2017communication}, a prevailing FL approach, averages models from all sites by their sample weights on the server side and then broadcasts the averaged result back to each site. Although FL has recently made promising progress in the medical image segmentation realm, most efforts still focus on improving the generalizability of a single global model via enhancing the aggregation strategy \cite{li2020federated,li2019privacy,liu2022ms} or employing data augmentation \cite{liu2021feddg,shen2021multi}. Moreover, almost all existing FL studies are conducted under fully supervised paradigms.

\begin{figure}[H]
    \vspace{-0.26cm}
    \centering
    \includegraphics[width=1\textwidth]{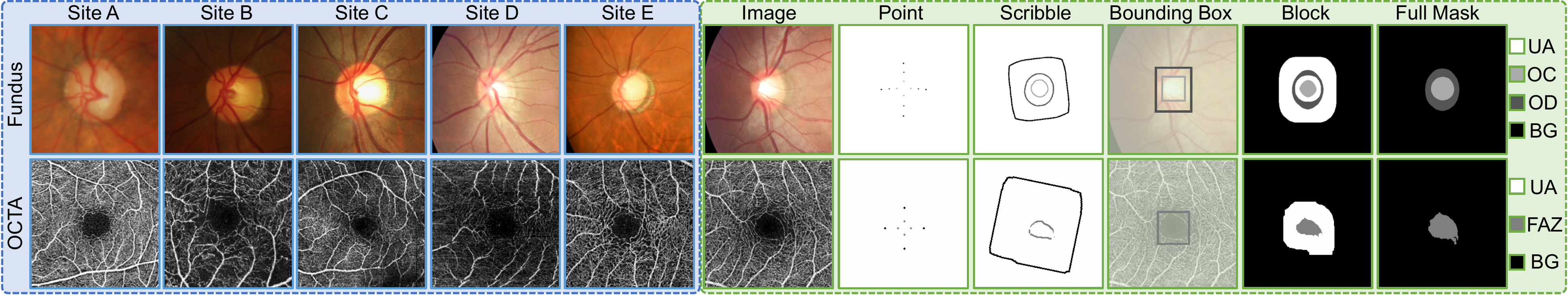}
    \vspace{-0.6cm}
    \caption{Left: Data samples from different sites showcase the domain shifts in their distributions; Right: Examples of various types of sparse annotations and the corresponding full masks. UA, OC, OD, FAZ, and BG respectively represent unlabeled area, optic cup, optic disc, foveal avascular zone, and background.} \label{fig1}
    \vspace{-0.5cm}
\end{figure}

A key obstacle in FL is statistical heterogeneity, mainly induced by non independent and identically distributed (non-IID) data, which makes it difficult for a single global model to perform well across all sites. FL's degradation could be even more severe in medical image analysis scenarios, as there may exist more diverse image domain shifts caused by differences in the imaging devices, protocols, patient populations, and physician expertise, as shown in Fig.~\ref{fig1} (Left). Personalized FL (pFL) has been proposed as a solution to these challenges, with the aim of training more customized local models for different sites. Existing pFL methods can be mainly classified into three categories: (1) fine-tuning the trained global model at each site, such as FedAvg with fine-tuning (FT) \cite{wang2019federated}; (2) partitioning the model into a global part and a personalized part, aggregating the global part on the server while retaining the locally trained parameters for the personalized part, such as FedBN and FedRep \cite{collins2021exploiting,li2021fedbn}; (3) combining information from other sites through local aggregation or knowledge distillation, such as FedAP and MetaFed \cite{chen2022metafed,lu2022personalized,zhang2022fedala}. Methods in category (1) risk forgetting common knowledge and overfitting local data, while those in (2) fail to leverage useful knowledge from other sites. Methods in category (3) relieve these issues, but local aggregation is typically performed at the model or layer level, which may result in excessive personalization and lead to performance degradation when the data heterogeneity is not that significant.

Recently, weakly supervised segmentation (WSS) has received significant research interest due to its ability to reduce annotation costs. WSS methods typically leverage sparsely-grained (e.g., point-, bounding box-, scribble-, block-wise) supervision through novel loss function designing, consistency learning, adversarial learning, or data synthesis \cite{liang2022tree,lin2022yolocurvseg,obukhov2019gated}. Integrating WSS into FL can further reduce the annotation cost at each site while exploiting the benefits of FL. However, there is currently a lack of research on WSS in the context of pFL. A more challenging yet more practical setting is different sites have different forms of weak labels, as shown in Fig.~\ref{fig1} (Right), which further introduces label heterogeneity. In such context, designing a unified and easily deployable pFL framework for the WSS setting is highly desirable.

Here, we propose a novel pFL framework for medical image segmentation, named FedICRA, which leverages heterogeneous weak supervision via adapt\textbf{I}ve \textbf{C}ontrastive \textbf{R}epresentation and \textbf{A}ggregation. Specifically, to facilitate model personalization, a channel selection based \textbf{S}ite \textbf{C}ontrastive \textbf{R}epresentation (\textbf{SC\\R}) module is applied, adaptively characterizing the data distributions of different sites as being tightly self-related and different from each other. An \textbf{A}daptive \textbf{A}ggregation (\textbf{AA}) module is proposed to element-wisely aggregate each local model with the global one towards the local objective at each site. Moreover, a weakly supervised loss is designed to uniformly leverage heterogeneous labels.

Our main contributions are four-fold: (1) To our knowledge, we are the first to propose a pFL framework for heterogeneous WSS. (2) \textbf{SCR} enhances inter-site representation contrast via channel attention, while \textbf{AA} element-wisely aggregates the global model's generic knowledge and each local model's specific knowledge. Both modules effectively promote personalization but also alleviate over-personalization. (3) A WSS objective that leverages a multiscale tree energy loss and a gated CRF loss is proposed to uniformly leverage heterogeneous labels and generate high-quality pseudo-labels for better training. (4) We validate the effectiveness of our proposed method on two medical image segmentation tasks and our approach achieves state-of-the-art (SOTA) performance on both tasks.
\section{Methodology}

\begin{figure}[t]
    \vspace{-0.26cm}
    \centering
    \includegraphics[width=\textwidth]{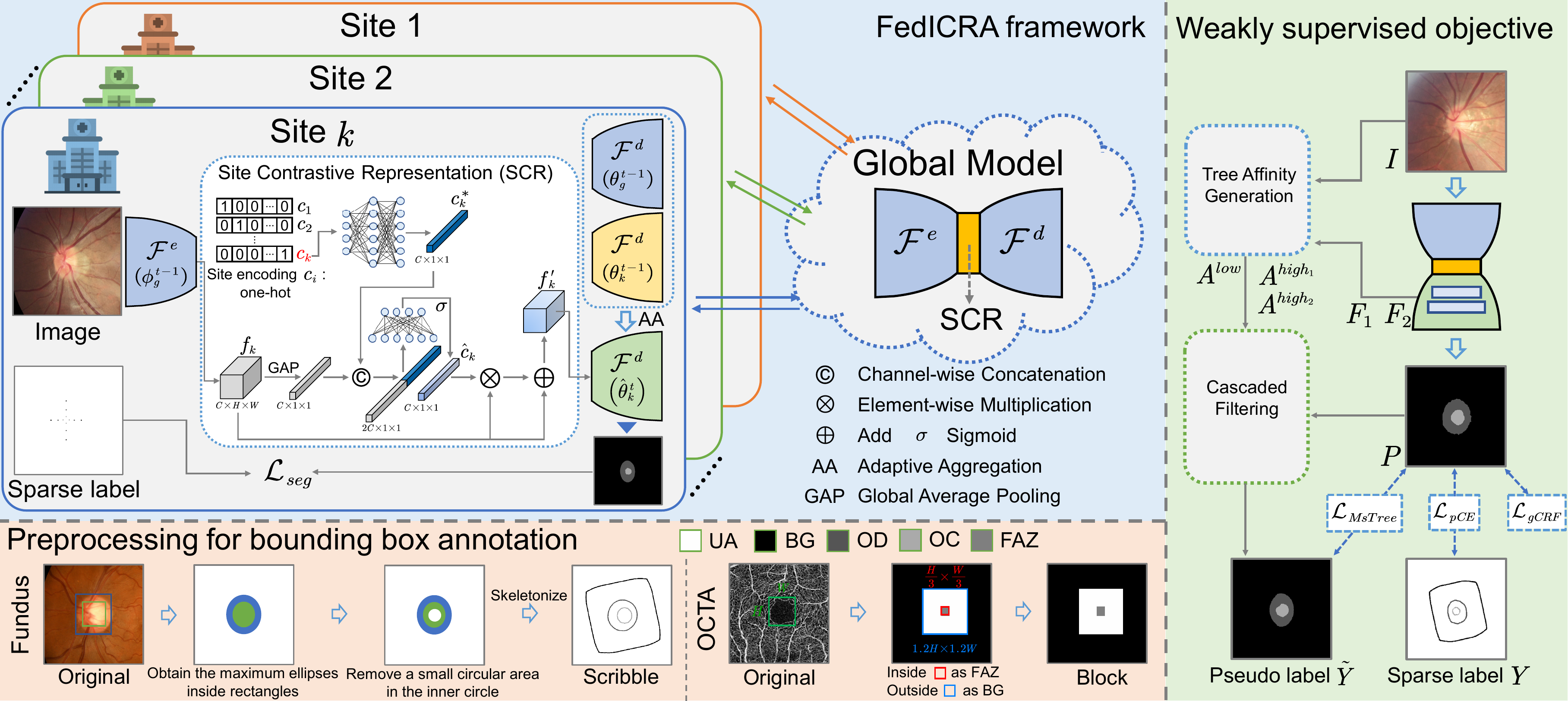}
    \vspace{-0.55cm}
    \caption{Schematic representation of the proposed FedICRA framework. {\color{white} and aaaa}} 
    \label{fig2}
    \vspace{-0.5cm}
\end{figure}

The overall pipeline of our FedICRA is provided in Fig.~\ref{fig2}. We first overview our pFL paradigm, and then introduce details of the \textbf{SCR} and \textbf{AA} modules, as well as our weakly supervised objective in the following paragraphs.

\textbf{Federation Paradigm Overview.} Suppose there are $K$ sites with private training data $D_1$, ..., $D_k$. These data are of distinct distributions and have different label types. With the help of a central server, FedICRA aims to collaboratively train a personalized model for each site without sharing data. The global and all local models share the same network architecture, namely UNet \cite{ronneberger2015u} with four encoding and decoding levels and an SCR module inserted in the middle. For more generalizable representations, we set the representation part of the network (encoder and SCR with parameter $\phi$) as the global part and the task head (decoder with parameter $\theta$) as the personalized part.
FedICRA alternatively updates between local sites and the global server in each communication round. At round $t$, all sites receive the same parameters ($\phi^{t-1}_g$,$\theta^{t-1}_g$) from the server. The global part is initialized with $\phi^{t-1}_g$, and the personalized part is initialized with $\hat{\theta}^{t}_k$ which is obtained from the AA module based on $\theta^{t-1}_g$ and the local parameters from the previous round (e.g., $\theta^{t-1}_k$ for site $k$). Each site updates its model by optimizing the local objective with its own data and its site encoding $c_k$ utilized in SCR
\vspace{-0.05cm}
\begin{equation}
    \phi_k^{t}, \theta_k^t \leftarrow \operatorname{GRD}\left(\phi_g^{t-1},\hat{\theta}^{t}_k \mid D_k, c_k\right),
\end{equation}
where $\operatorname{GRD}(\cdot)$ denotes the local gradient-based update. As for the aggregation of the global model, we adopt the weighted averaging strategy in FedAvg \cite{mcmahan2017communication}.

\textbf{Site Contrastive based Channel Selection.} Personalized FL paradigms may suffer from confusion or over-personalization when data heterogeneity is low, performing even worse than traditional FL methods \cite{zhang2022fedala}. 
Hence, the SCR module is designed to enhance the distance/contrast of inter-site data representations through site-contrastive learning based channel attention, which in turn facilitates personalization. Specifically, taking the $k$-th site as an example, a one-hot site encoding $c_k$ (i.e., the $k$-th position is 1 and others are 0) and the output feature $f_k$ from the encoder $\mathcal{F} ^e$ are given. $c_k$ is expanded to a length of $C$ through two fully connected layers to obtain $c_k^*$, which is then concatenated with the global average pooled feature of $f_k$. After passing through a fully connected layer with Sigmoid activation, the site channel attention value $\hat{c}_k$ is obtained. The final feature $f_{k}^{'}$ is obtained in a residual manner: $f_{k}^{'} = f_k +  f_k \otimes \hat{c}_k$, where $\otimes$ denotes element-wise multiplication. The final feature $f_{k}^{'}$ is then fed into the decoder $\mathcal{F} ^d$ to obtain the final segmentation. Note that during training, we assign different site encodings $\left\{ {c}_i \right\} _{i=1}^{K}$ sequentially to obtain $\left\{ \hat{c}_i \right\} _{i=1}^{K}$ with different site styles/distributions. The contrastive objective used to increase the inter-site embedding difference is formulated as
\begin{equation}
    \mathcal{L}_{con}=-\frac{1}{K-1} \sum_{i=1}^K\left|\hat{c}_k-\operatorname{StopGradient}\left(\hat{c}_i\right)\right| \text {, s.t. } i \neq k. 
\end{equation}

\textbf{Adaptive Head Aggregation.} Following existing pFL paradigms, we globally share the representation part of the segmentation network (encoder $\mathcal{F} ^e$ and the SCR module) and personalize the task head (decoder $\mathcal{F} ^d$). To element-wisely aggregate two models without introducing multiple aggregation weight matrices, we adopt an adaptive learning based approach similar to residual learning for updating the weight matrices. The adaptive head aggregation is formulated as
\begin{equation}
    \vspace{-0.01cm}
    \hat{\theta}_i^t:=\theta_i^{t-1}+\left(\theta^{t-1}_g-\theta_i^{t-1}\right) \odot W_i,
    \vspace{-0.01cm}
\end{equation}
where $W_i$ is a learnable weight matrix. Inspired by \cite{luo2019adaptive}, we clip and make $w \in[0,1]$, $\forall w \in W_i$ for regularization via $\sigma(w)=\max (0, \min (1, w))$. At the beginning, every element in $W_i$ is initialized to be 1, and then iteratively gets updated via
\begin{equation}
    \vspace{-0.03cm}
    W_i \leftarrow \operatorname{GRD}\left(W_i^{t-1} \mid D_i, \hat{\theta}_i^t\right).
    \vspace{-0.01cm}
\end{equation}
In the update process of $W_i$, Eqs. (3) and (4) are respectively utilized to alternatively update $\hat{\theta}_i^t$ and $W_i$. Upon convergence, the initialization parameter $\hat{\theta}_i^t$ for $\mathcal{F} ^d$ in each federation round is obtained, after which local models are trained using Eq. (1). Note that when $t>2$, only one epoch is trained to obtain $W_i$.

\textbf{Weakly Supervised Objective.} The WSS task aims to train a dense prediction model based on sparse annotations, wherein most pixels are unlabeled. To reduce the communication cost and enable all sites to cooperatively utilize different sparse labels in a uniform and easily deployable manner, we boost the WSS performance by optimizing the objective function rather than the network architecture nor the training strategy. For each image in the training set, it can be divided into a labeled set $I_L$ and an unlabeled set $I_U$. The commonly used partial cross-entropy loss ($\mathcal{L} _{pCE}$) \cite{tang2018normalized} is employed for $I_L$, which is formulated as
\begin{equation}
    \vspace{-0.01cm}
    \mathcal{L}_{pCE}=-\frac{1}{\left|I_L\right|} \sum_{\forall i \in I_L} Y_i \log \left(P_i\right),
    \vspace{-0.01cm}
\end{equation}
where $Y_i$ and $P_i$ respectively denote the sparse ground truth and the predicted probability of pixel $i$. The key to improving the WSS performance lies in how to effectively utilize $I_U$ to generate accurate pseudo-labels. Since pixels belonging to the same object share similar patterns across different feature levels, we follow the design of tree filters \cite{liang2022tree,song2019learnable}; we adapt the tree energy loss into a multi-scale recursive version that better accommodates medical image segmentation. Concretely, triple affinity matrices $A^{low}$, $A^{high_1}$ and $A^{high_2}$ are calculated via a \textit{Borůvka} algorithm \cite{gallager1983distributed} based generation module respectively from the original image as well as feature maps from the second and third decoding stages. $A^{low}$ contains object boundary information, while $A^{high_1}$ and $A^{high_2}$ maintain semantic consistency. By employing the same cascaded filtering operation $\mathcal{F}(\cdot)$ in \cite{liang2022tree}, the pseudo-labels $\tilde{Y}$ can be generated through
\begin{equation}
    \vspace{-0.01cm}
    \tilde{Y}=\mathcal{F}\left(\mathcal{F}\left(\mathcal{F}\left(P, A^{\text {low}}\right), A^{{high}_1}\right),A^{{high}_2} \right),
    \vspace{-0.01cm}
\end{equation}
and the multi-scale recursive tree energy loss ($\mathcal{L}_{MsTree}$) goes as
\begin{equation}
    \vspace{-0.01cm}
    \mathcal{L}_{\text {MsTree}}=-\frac{1}{\left|I_U\right|} \sum_{\forall i \in I_U} \left|P_i-\tilde{Y}_i\right|.
    \vspace{-0.01cm}
\end{equation}
Additionally, the gated CRF loss ($\mathcal{L}_{gCRF}$) \cite{obukhov2019gated} is employed as an extra regularization term to further increase the edge accuracy and reduce outliers in $P$ and $\tilde{Y}$. Therefore, with trade-off parameters $\lambda_1, \lambda_2, \lambda_3$, the WSS objective and the total local objective can be respectively expressed as
\begin{equation}
    \mathcal{L}_{seg}=\mathcal{L}_{pCE}+\lambda_1\mathcal{L}_{\text {MsTree}}+\lambda_2\mathcal{L}_{gCRF},
\end{equation}
\begin{equation}
    \mathcal{L}_i=\mathcal{L}_{seg}+\lambda_3\mathcal{L}_{con}.
\end{equation}

\section{Experiments and Results}
\vspace{-0.05cm}
\textbf{Datasets and Preprocessing.} Extensive experiments are conducted to verify the effectiveness of the proposed FedICRA on two medical image segmentation datasets, including optic disc/cup (ODOC) segmentation from fundus images and foveal avascular zone (FAZ) segmentation from optical coherence tomography angiography (OCTA) images. Both fundus images \cite{sivaswamy2015comprehensive,fumero2011rim,orlando2020refuge,wu2022gamma} and OCTA images \cite{agarwal2020foveal,li2020ipn,wang2021deep,ma2020rose} are obtained from publicly available datasets with heterogeneous distributions. For fundus images, we follow \cite{liu2021feddg} to create the first four sites and add GAMMA \cite{wu2022gamma} as the fifth site. The datasets for the five sites consist of \{101, 159, 400, 400, 200\} samples. For OCTA images, we partition the original dataset into five sites, respectively containing \{304, 200, 300, 1012, 39\} samples. Fundus images are center-cropped and then resized to 384 $\times$ 384 while OCTA images are directly resized to 256 $\times$ 256. We utilize automated algorithms to simulate point, scribble (two styles), bounding box, and block annotations for each site based on the original segmentation masks. The train-test splits of all datasets are in line with the original dataset partitioning; more information about the constructed datasets can be found in Table {\color{blue} A1} in the appendix.

Data preprocessing includes normalizing all image intensities to between 0 and 1, while data augmentation includes randomly flipping images horizontally and vertically as well as rotation (spanning from -45° to 45°). Moreover, since bounding box annotations are not a direct sparse supervision signal in WSS, we perform certain preprocessing to covert them into scribble and block respectively for ODOC and FAZ segmentation, as shown in the lower left panel of Fig. \ref{fig2}.

\vspace{-0.5cm}
\subsubsection*{Implementation Details.} 
All compared FL methods and the proposed FedICRA are implemented with PyTorch using NVIDIA A100 GPUs. We employ the vanilla UNet as the model architecture, with the number of channels progressively increasing from 16 to 256 from top to bottom. We use the AdamW optimizer with an initial learning rate of 1 $\times$ 10$^{-2}$ to optimize the parameters and the polynomial policy with $power$ = 0.9 to dynamically adjust the learning rate \cite{mishra2019polynomial}. The trade-off coefficients $\lambda_1, \lambda_2, \lambda_3$ and batch size are respectively set to be 0.1, 0.1, 1 and 12. We train all FL methods for 500 federation rounds to ensure fair performance comparisons.

\begin{figure}[t]
    \vspace{-0.26cm}
    \centering
    \includegraphics[width=1\textwidth]{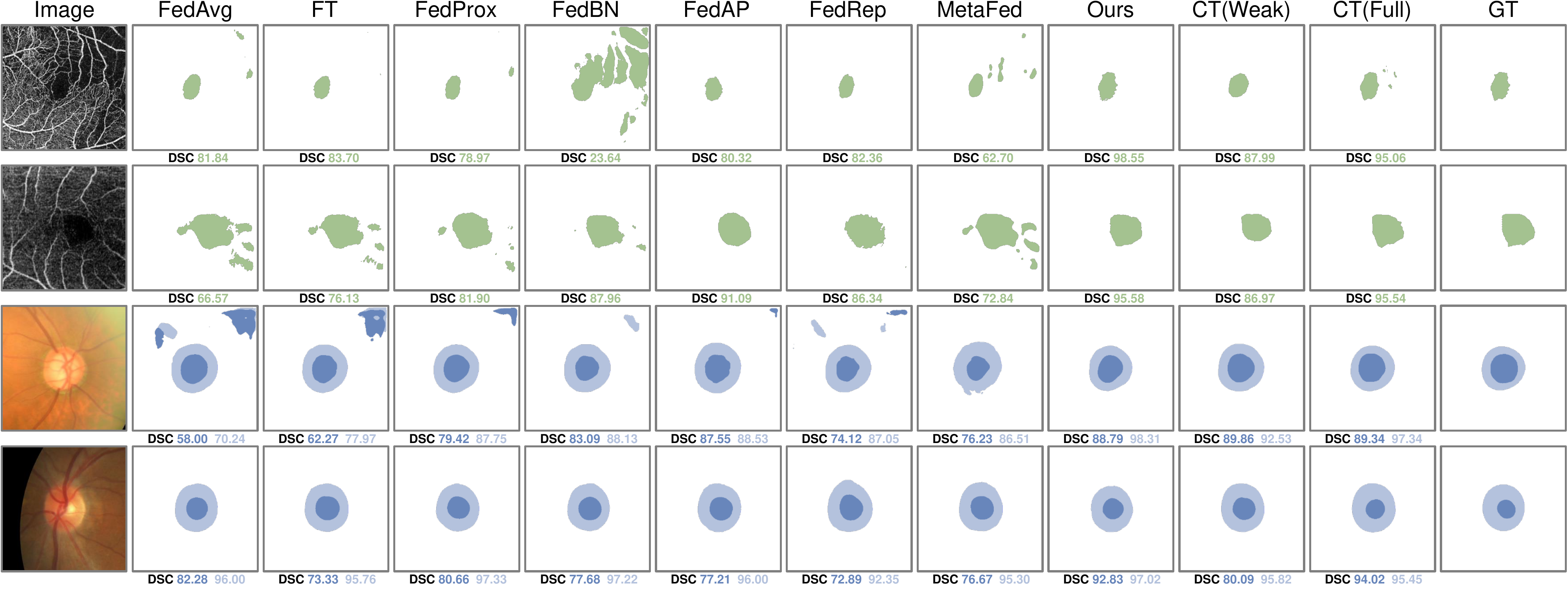}
    \vspace{-0.65cm}
    \caption{Visualization results from FedICRA and other SOTA methods (\textbf{CT} indicates centralized training, with \textbf{Weak} and \textbf{Full} respectively denoting utilizing sparse annotations and full masks).} \label{fig3}
    \vspace{-0.5cm}
\end{figure}

\begin{table}[!h]
    \vspace{-0.6cm}
    \caption{
        Performance comparisons (DSC) of different FL methods as well as different local and centralized training settings on ODOC segmentation. The best results are highlighted in bold and the second-best ones are underlined.}
    \label{table1}
    \setlength{\tabcolsep}{0.6mm}
    \vspace{+0.1cm}
    \centering
    \resizebox{\textwidth}{!}{
        \begin{tabular}{l|ccccc|c|ccccc|c|c}
            \specialrule{0.12em}{0pt}{0pt}
            & \multicolumn{6}{c|}{DSC (OD) $\uparrow$}                                                    & \multicolumn{6}{c|}{DSC (OC) $\uparrow$}                                                     & \multirow{2}{*}{Overall}        \\
            \cline{1-13}
            Methods & Site A          & Site B           & Site C          & Site D           & Site E           & Avg.  & Site A           & Site B           & Site C          & Site D           & Site E           & Avg.  &  \\
\specialrule{0.05em}{0pt}{0pt}
FedAvg \cite{mcmahan2017communication}        & 92.58±5.69 & 83.90±16.18 & 94.67±2.20 & 93.36±6.71  & 88.92±8.20  & 90.68 & 83.16±14.61 & 69.91±18.68 & \underline{85.35±7.25} & 82.80±11.32 & 86.76±12.79 & 81.60 & 86.14   \\
FedProx \cite{li2020federated}        & 95.19±4.81 & 82.14±18.66 & \underline{95.03±2.45} & 88.28±11.84 & \underline{91.06±8.92}  & 90.34 & 82.18±14.42 & 70.14±17.98 & 84.64±7.05 & 81.33±12.79 & \underline{87.19±9.63}  & 81.10 & 85.72   \\
FT \cite{wang2019federated}             & \underline{95.97±2.17} & 91.09±3.62  & 94.83±2.18 & 93.34±3.49  & 90.10±8.19  & 93.06 & \underline{85.02±13.79} & 80.76±10.89 & 82.51±7.84 & 82.95±7.50  & 86.88±11.16 & \underline{83.62} & 88.34   \\
FedBN \cite{li2021fedbn}          & 95.88±1.94 & \underline{92.66±3.32}  & 94.80±2.64 & \underline{95.06±2.73}  & 90.72±10.28 & 93.82 & 84.54±12.93 & \textbf{82.09±10.22} & 83.79±7.79 & 86.48±6.26  & 79.38±15.28 & 83.25 & 88.54   \\
FedAP \cite{lu2022personalized}          & 95.81±2.17 & 92.24±3.31  & \textbf{95.45±1.91} & \textbf{95.17±2.26}  & 90.73±10.36 & \underline{93.88} & 83.97±13.78 & 79.89±12.41 & 83.55±7.20 & \underline{86.78±6.86}  & 83.47±12.86 & 83.53 & \underline{88.71}   \\
FedRep \cite{collins2021exploiting}         & 95.28±2.70 & 88.10±8.98  & 92.68±4.34 & 92.34±5.91  & 88.75±10.91 & 91.43 & 82.47±14.19 & 76.22±14.68 & 82.14±8.10 & 83.04±8.49  & 81.29±13.64 & 81.03 & 86.23   \\
MetaFed \cite{chen2022metafed}      & 91.71±8.77 & 91.48±7.55  & 88.70±9.20 & 91.76±6.96  & 82.29±11.94 & 89.19 & 80.10±18.55 & 79.43±11.13 & 83.14±8.02 & 85.95±6.87  & 81.38±15.27 & 82.01 & 85.59   \\
FedICRA (Ours) & \textbf{96.51±1.52} & \textbf{93.85±2.74}  & 94.62±1.79 & 94.06±2.68  & \textbf{94.10±4.23}  & \textbf{94.63} & \textbf{85.44±13.32} & \underline{81.42±10.36} & \textbf{87.89±5.51} & \textbf{88.76±5.59}  & \textbf{89.68±5.30} & \textbf{86.64} & \textbf{90.63}   \\
\specialrule{0.05em}{0pt}{0pt}
LT (Weak)\cellcolor{gray!30}      & 94.83±4.25\cellcolor{gray!30}& 90.65±5.16\cellcolor{gray!30} & 89.31±8.68\cellcolor{gray!30} & 76.38±15.11\cellcolor{gray!30} & 86.11±14.03\cellcolor{gray!30} & 87.45\cellcolor{gray!30} & 84.50±12.50\cellcolor{gray!30} & 79.31±14.33\cellcolor{gray!30} & 84.11±7.70\cellcolor{gray!30} & 77.02±9.68\cellcolor{gray!30}  & 81.38±16.29\cellcolor{gray!30} & 81.26\cellcolor{gray!30} & 84.36\cellcolor{gray!30}   \\
CT (Weak)\cellcolor{gray!30}       & 95.45±2.56\cellcolor{gray!30}  & 91.41±3.32\cellcolor{gray!30}   & 95.45±1.94\cellcolor{gray!30}  & 91.92±4.91\cellcolor{gray!30}   & 91.21±8.55\cellcolor{gray!30}   & 93.09\cellcolor{gray!30}  & 84.25±13.95\cellcolor{gray!30}  & 78.94±11.40\cellcolor{gray!30}  & 84.63±6.49\cellcolor{gray!30}  & 85.68±9.04\cellcolor{gray!30}   & 87.08±12.14\cellcolor{gray!30}  & 84.12\cellcolor{gray!30}  & 88.60\cellcolor{gray!30}     \\
LT (Full)\cellcolor{gray!30}      & 96.28±2.69\cellcolor{gray!30}  & 96.21±2.50\cellcolor{gray!30}   & 95.21±3.95\cellcolor{gray!30}  & 95.56±1.89\cellcolor{gray!30}   & 94.33±10.64\cellcolor{gray!30}  & 95.52\cellcolor{gray!30}  & 87.94±8.08\cellcolor{gray!30}   & 81.72±12.28\cellcolor{gray!30}  & 87.22±6.41\cellcolor{gray!30}  & 88.60±7.37\cellcolor{gray!30}   & 86.53±12.24\cellcolor{gray!30}  & 86.40\cellcolor{gray!30}   & 90.96\cellcolor{gray!30}    \\
CT (Full)\cellcolor{gray!30}      & 96.97±1.37\cellcolor{gray!30}  & 94.28±4.25\cellcolor{gray!30}   & 96.06±1.77\cellcolor{gray!30}  & 96.20±1.66 \cellcolor{gray!30}  & 95.52±8.37\cellcolor{gray!30}   & 95.81\cellcolor{gray!30}  & 87.40±11.90\cellcolor{gray!30}  & 82.90±9.07\cellcolor{gray!30}   & 88.30±6.39\cellcolor{gray!30} & 89.17±5.82\cellcolor{gray!30}   & 89.25±11.47\cellcolor{gray!30}  & 87.40\cellcolor{gray!30}  & 91.61\cellcolor{gray!30}   \\
\specialrule{0.12em}{0pt}{0pt}

\end{tabular}
    }
\vspace{-0.4cm}
\end{table}

\vspace{-0.5cm}
\subsubsection*{Evaluation Results.} We compare FedICRA with several FL frameworks including traditional FL (FedAvg \cite{mcmahan2017communication} and FedProx \cite{li2020federated}) and SOTA pFL methods (i.e., FT \cite{wang2019federated}, FedBN \cite{li2021fedbn}, FedAP \cite{lu2022personalized}, FedRep \cite{collins2021exploiting} and MetaFed \cite{chen2022metafed}). Most of these methods are originally designed for classification, and we endeavor to maintain their design principles while adapting them to segmentation tasks. We also compare with some baseline and ideal settings, including local training (LT) with weak labels and full labels, as well as centralized training (CT). All methods are evaluated using two metrics, i.e., the Dice similarity coefficient (DSC) and the 95\% Hausdorff distance (HD95[px]).

Table \ref{table1} and Table {\color{blue} A2} (in the appendix) display the quantitative results for the ODOC segmentation task. Compared to LT, all FL methods improve the overall segmentation performance of all sites. Sites C and D achieve significant performance improvement after participating in FL as their local data are distributed diversely, making it difficult for them to train sufficiently powerful models using local data alone. Due to the high data heterogeneity across sites, as shown in Fig. {\color{blue} A2}, pFL methods benefit from training personalized models for different sites and generally achieve higher performance than centralized FL methods. FedAP uses BN layers to measure the inter-site data distribution similarity and aggregate personalized models, thereby extracting useful information from other sites and achieving good performance. Notably, our FedICRA achieves further performance improvement with a DSC about 2\% higher than FedAP, approaching centralized training with fully supervised labels.

\begin{table}[t]
    \vspace{-0.65cm}
    \caption{
        Performance comparisons (DSC and HD95) of different FL methods as well as different local and centralized training settings on FAZ segmentation.}
    \label{table2}
    \setlength{\tabcolsep}{0.6mm}
    \vspace{+0.1cm}
    \centering
    \resizebox{\textwidth}{!}{
        \begin{tabular}{l|ccccc|c|ccccc|c}
            \specialrule{0.12em}{0pt}{0pt}
            & \multicolumn{6}{c|}{DSC$\uparrow$}                                                  & \multicolumn{6}{c}{HD95[px]$\downarrow$}                                                    \\ \hline
        Methods & Site A           & Site B          & Site C           & Site D          & Site E           & Avg.  & Site A           & Site B           & Site C           & Site D           & Site E           & Avg.  \\
\specialrule{0.05em}{0pt}{0pt}
FedAvg \cite{mcmahan2017communication}        & 76.56±28.93 & 90.65±6.01 & 77.64±15.72 & 88.85±6.13 & 86.55±6.13  & 84.05 & \underline{5.85±6.56}   & 8.28±6.12   & 22.73±38.04 & 12.45±15.87 & 21.59±35.61 & 14.18 \\
FedProx \cite{li2020federated}        & 73.87±29.34 & 89.70±7.19 & 77.44±18.97 & 89.15±5.24 & 83.78±7.75  & 82.79 & 7.38±8.83   & 8.63±5.83   & 23.07±39.18 & 12.18±15.51 & 22.62±35.25 & 14.78 \\
FT \cite{wang2019federated}             & \underline{83.18±7.91}  & \underline{91.96±3.89} & 78.58±16.28 & 88.75±5.54 & \underline{88.10±7.40}  & \underline{86.11} & 10.11±13.02 & 8.86±11.00  & 20.84±36.52 & 11.53±13.30 & 20.17±36.16 & 14.3  \\
FedBN \cite{li2021fedbn}          & 62.28±25.01 & 90.11±5.17 & 60.93±21.41 & 87.87±5.95 & 53.05±23.92 & 70.84 & 51.03±56.30 & 9.42±10.89  & 60.14±58.95 & 11.84±12.11 & 20.31±8.85  & 30.54 \\
FedAP \cite{lu2022personalized}          & 62.23±10.57 & 87.82±4.39 & 72.67±14.79 & 90.36±5.23 & 66.00±9.47  & 75.82 & 18.50±15.87 & \underline{7.71±2.54}   & \underline{14.51±20.16} & 8.30±5.71   & \underline{18.60±4.70}  & \underline{13.52} \\
FedRep \cite{collins2021exploiting}         & 78.55±18.86 & 91.69±4.40 & \underline{79.29±18.13} & \underline{91.04±4.95} & 86.83±11.46 & 85.48 & 16.07±22.16 & 9.14±16.50  & 15.24±26.29 & \underline{7.64±3.75}   & 19.92±36.11 & 13.60  \\
MetaFed \cite{chen2022metafed}        & 70.99±29.03 & 86.22±9.19 & 74.15±21.40 & 88.77±6.24 & 74.62±21.15 & 78.95 & 9.18±12.04  & 10.37±6.43  & 24.61±38.99 & 13.91±16.95 & 39.61±47.34 & 19.54 \\
FedICRA (Ours) & \textbf{88.59±6.40}  & \textbf{97.42±1.52} & \textbf{90.84±8.72}  & \textbf{94.32±4.73} & \textbf{95.28±0.99}  & \textbf{93.29} & \textbf{5.77±3.14}   & \textbf{4.05±2.81}   & \textbf{8.24±13.35}  & \textbf{5.75±3.68}   & \textbf{4.14±1.04}   & \textbf{5.59}  \\
\specialrule{0.05em}{0pt}{0pt}
LT (Weak)\cellcolor{gray!30}      & 73.74±27.11\cellcolor{gray!30} & 91.90±2.77\cellcolor{gray!30} & 79.03±16.57\cellcolor{gray!30} & 85.78±7.32\cellcolor{gray!30} & 79.89±19.92\cellcolor{gray!30} & 82.07\cellcolor{gray!30} & 30.94±47.78\cellcolor{gray!30} & 42.49±45.28\cellcolor{gray!30} & 20.81±35.48\cellcolor{gray!30} & 10.87±4.81\cellcolor{gray!30}  & 72.08±56.19\cellcolor{gray!30} & 35.44\cellcolor{gray!30} \\
CT (Weak)\cellcolor{gray!30}      & 74.93±23.48\cellcolor{gray!30} & 89.24±3.94\cellcolor{gray!30} & 76.64±16.37\cellcolor{gray!30} & 89.25±5.40\cellcolor{gray!30} & 86.50±4.71\cellcolor{gray!30}  & 83.31\cellcolor{gray!30} & 8.38±6.28\cellcolor{gray!30}   & 7.54±2.42\cellcolor{gray!30}   & 10.46±10.96\cellcolor{gray!30} & 8.87±3.78\cellcolor{gray!30}   & 8.28±1.41\cellcolor{gray!30}   & 8.7\cellcolor{gray!30}   \\
LT (Full)\cellcolor{gray!30}      & 90.88±5.06\cellcolor{gray!30}  & 97.75±2.13\cellcolor{gray!30} & 89.22±17.63\cellcolor{gray!30} & 95.14±5.46\cellcolor{gray!30} & 95.11±3.00\cellcolor{gray!30}  & 93.62\cellcolor{gray!30} & 5.31±6.00\cellcolor{gray!30}   & 3.36±3.49\cellcolor{gray!30}   & 7.84±13.20\cellcolor{gray!30}  & 4.80±3.08\cellcolor{gray!30}   & 4.54±2.61\cellcolor{gray!30}   & 5.17\cellcolor{gray!30}  \\
CT (Full)\cellcolor{gray!30}      & 90.93±0.052\cellcolor{gray!30} & 97.23±1.42\cellcolor{gray!30} & 91.49±8.26\cellcolor{gray!30}  & 94.88±4.27\cellcolor{gray!30} & 95.38±1.17\cellcolor{gray!30}  & 93.98\cellcolor{gray!30} & 4.57±3.04\cellcolor{gray!30}   & 3.66±3.33\cellcolor{gray!30}   & 6.86±10.32\cellcolor{gray!30}  & 4.82±3.17\cellcolor{gray!30}   & 2.99±0.86\cellcolor{gray!30}   & 4.58\cellcolor{gray!30} \\
    \specialrule{0.12em}{0pt}{0pt}

\end{tabular}
    }
\vspace{-0.3cm}
\end{table}

\begin{table}[]
    \begin{minipage}{0.45\linewidth}
        \vspace{-0.95cm}
        \caption{
            Ablation study on the four key elements in FedICRA (DSC).}
        \label{table3}
        \setlength{\tabcolsep}{0.5mm}
        \vspace{+0.15cm}
        \centering
        \resizebox{0.97\textwidth}{!}{
        \begin{tabular}{cccc|cc|c}
        \specialrule{0.12em}{0pt}{0pt}
        \textbf{AA} & \textbf{SCR} & $\mathcal{L}_{MsTree}$ & $\mathcal{L}_{gCRF}$ & OD             & OC             & FAZ            \\
        \specialrule{0.05em}{0pt}{0pt}
    
        -           & -            & -             & -                    & 90.68          & 81.60           & 84.05          \\
        \checkmark           &              &               &                      & 92.98          & 84.67          & 88.96          \\
                    & \checkmark            &               &                      & 93.19          & 84.07          & 88.27          \\
                    \checkmark           & \checkmark            &               &                      & 93.65          & 85.10           & 89.20           \\
        \specialrule{0.05em}{0pt}{0pt}
                    \checkmark           & \checkmark            & \checkmark             &                      & \underline{93.86}          & 85.57          & 89.38          \\
                    \checkmark           & \checkmark            &               & \checkmark                    & \textbf{94.63}          & \underline{85.82}          & \underline{92.47}          \\
                    \checkmark           & \checkmark           & \checkmark             & \checkmark                    & \textbf{94.63} & \textbf{86.64} & \textbf{93.29} \\
        \specialrule{0.12em}{0pt}{0pt}
                    
        \end{tabular}
        }
    \end{minipage}
    \vspace{-0.65cm}
    \begin{minipage}{0.55\linewidth}
        \centering
        \includegraphics[width=0.4\textwidth]{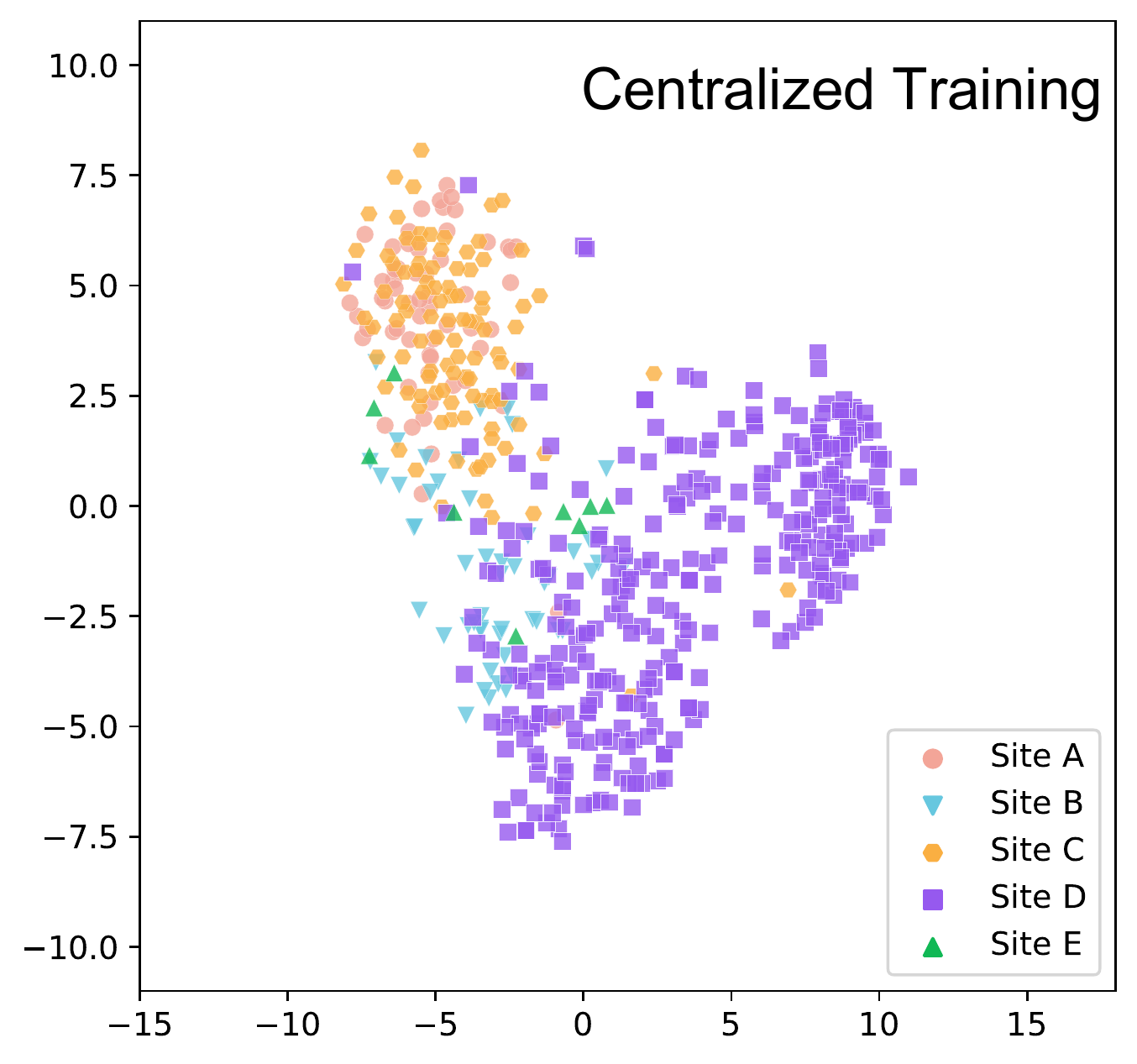}
        \hspace{+0.2cm}
        \includegraphics[width=0.4\textwidth]{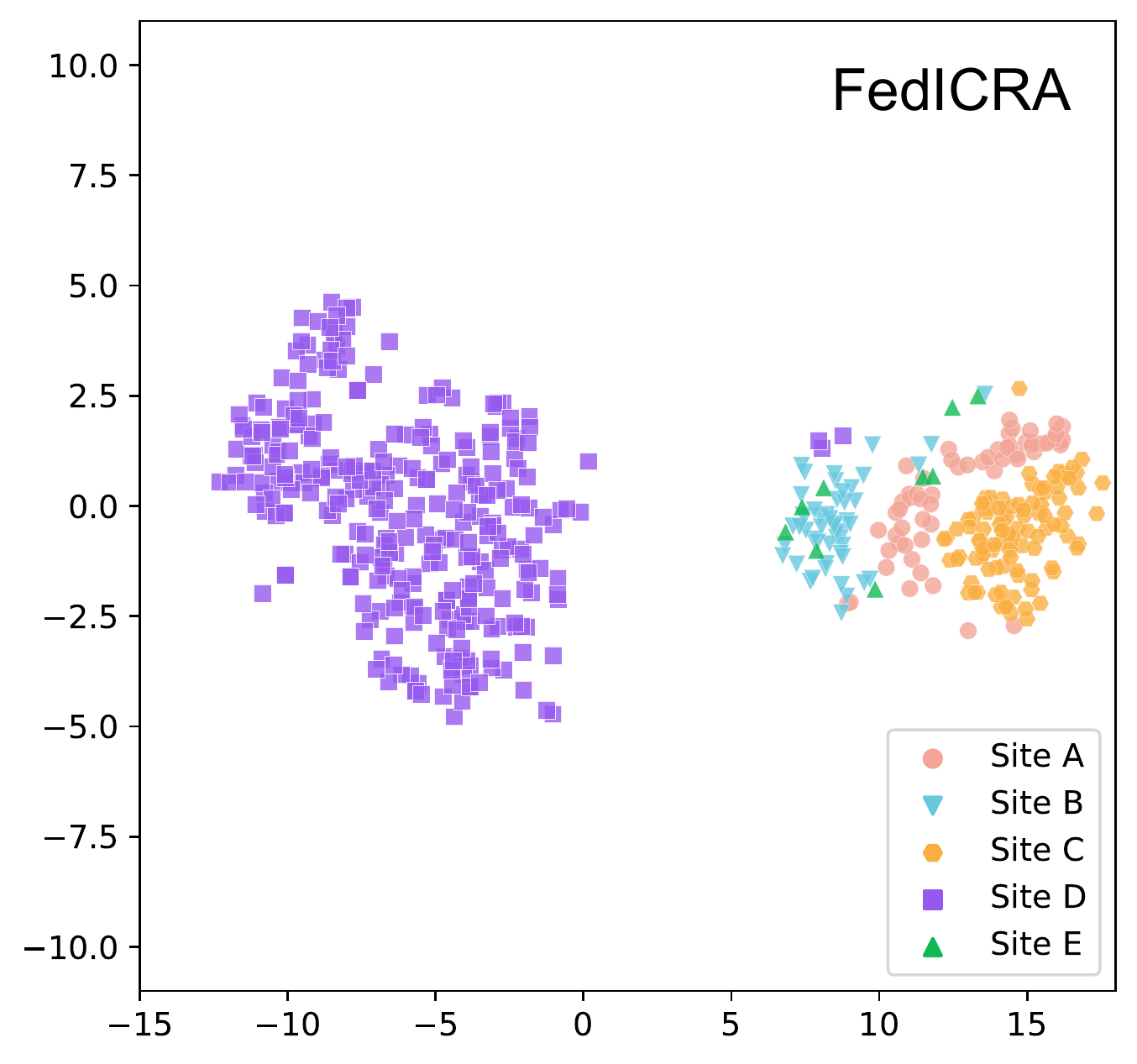}
        \vspace{+0.1cm}
        \captionof{figure}{Comparison of data embeddings on OCTA test sets via t-SNE visualization.} 
        \label{fig4} 
    
    \end{minipage}
    \end{table}

Table \ref{table2} tabulates the quantitative results for the FAZ segmentation task. Different from the ODOC task, due to low data heterogeneity as shown in the left part of Fig. \ref{fig4}, most pFL methods (such as FedBN, FedAP and MetaFed) encounter confusion or over-personalization, performing even worse than FedAvg. And FT attains the second-best performance. Our FedICRA leverages the SCR module to enhance the inter-site data representation contrast (Fig. \ref{fig4}), thereby mitigating the aforementioned issue. As a result, it is still capable of training high-performance personalized model for each site and achieves the best performance across all sites. FedICRA outperforms FT by a large margin and achieves performance close to CT with full supervision. Visualization results for the two tasks are provided in Fig. \ref{fig3} and Fig. {\color{blue} A3}.

We also calculate the p-values of the DSC values between FedICRA and the second-best method on OD, OC and FAZ tasks (respective 6.79 $\times$ 10$^{-6}$, 1.09 $\times$ 10$^{-18}$ and 1.63 $\times$ 10$^{-77}$), identifying statistically significant performance improvements of FedICRA. Table \ref{table3} shows the ablation results on the four key components (i.e., \textbf{AA}, \textbf{SCR}, $\mathcal{L}_{MsTree}$, and $\mathcal{L}_{gCRF}$) of FedICRA. The first row represents the performance of FedAvg. Experimental results demonstrate the importance of the four key components. Even without the two WSS losses, incorporating the two FL modules still outperforms other FL methods on both segmentation tasks. It is worth noting that the last two rows of Table \ref{table1} and Table \ref{table2} show that centerlizing data or using FL may not provide significant improvements over CT under the fully supervised setting. However, FL under weak supervision clearly highlights the importance of collaboration to reduce annotation costs and improve performance.
\section{Conclusion}
\vspace{-0.038cm}

This paper presents a novel pFL framework, FedICRA, for personalized medical image segmentation. Through extensive experiments on two different modalities and two different segmentation tasks, the effectiveness of the proposed method has been demonstrated. Additionally, we have shown that under heterogeneous weakly-supervised conditions, FedICRA can achieve segmentation performance that is comparable to that of labor-intensive fully-supervised centralized training. 

\section*{Acknowledgments}
This study was supported by the Shenzhen Basic Research Program (JCYJ20190\\809120205578); the National Natural Science Foundation of China (62071210); the Shenzhen Science and Technology Program (RCYX20210609103056042); the Shenzhen Basic Research Program (JCYJ20200925153847004); the Shenzhen Science and Technology Innovation Committee Program (KCXFZ202012211734\\0001).


\newpage
\section*{Appendix}
\setcounter{figure}{0}
\setcounter{table}{0}
\renewcommand{\thetable}{A\arabic{table}}
\renewcommand{\thefigure}{A\arabic{figure}}

\begin{table}[htbp]
    \caption{
        Details of the datasets utilized in our experiments. Scribble$^2$ refers to scribbles of a different style generated using another automated algorithm.}
    \label{tableA1}
    \setlength{\tabcolsep}{1mm}
    \vspace{+0.2cm}
    \resizebox{\textwidth}{!}{
    \begin{tabular}{ccccccc}
        \specialrule{0.12em}{0pt}{0.5pt}
        Modality                & Site & Original Dataset & \begin{tabular}[c]{@{}c@{}}Train-Test \\ Sample Size\end{tabular} & Annotation Type & \begin{tabular}[c]{@{}c@{}}Processed \\ Resolution {[}px{]}\end{tabular} & Manufactor                            \\
        \specialrule{0.05em}{0.5pt}{0.5pt}
        \multirow{5}{*}{Fundus} & A    & Drishti-G        & 50:51                                                             & Scribble         & \multirow{5}{*}{384×384}                                                 & (Aravind eye hospital)                \\
                                & B    & RIM-ONE-r        & 99:60                                                             & Scribble$^2$        &                                                                          & Nidek AFC-210                         \\
                                & C    & REFUGE-train     & 320:80                                                            & Bounding box     &                                                                          & Zeiss Visucam 500                     \\
                                & D    & REFUGE-val       & 320:80                                                            & Point            &                                                                          & Canon CR-2                            \\
                                & E    & GAMMA            & 100:100                                                           & Block            &                                                                          & KOWA camera \&  Topcon TRC-NW400      \\
        \specialrule{0.05em}{0.5pt}{0.5pt}
        \multirow{5}{*}{OCTA}   & A    & FAZID            & 244:60                                                            & Scribble         & \multirow{5}{*}{256×256}                                                 & Cirrus 5000 Angioplex                 \\
                                & B    & OCTA500-3M       & 150:50                                                            & Point            &                                                                          & RTVue-XR, Optovue, CA                 \\
                                & C    & OCTA500-6M       & 200:100                                                           & Block            &                                                                          & RTVue-XR, Optovue, CA                 \\
                                & D    & OCTA-25K (sOCTA-3x3) & 708:304                                                           & Bounding box     &                                                                          & Triton DRI-OCT, Topcon Inc            \\
                                & E    & ROSE             & 30:9                                                              & Scribble$^2$        &                                                                          & RTVueXR Avanti SD-OCT system, Optovue \\
                                \specialrule{0.12em}{0pt}{0pt}
                                
        \end{tabular}
    }
    \end{table}

\begin{figure}[H]
    \vspace{-0.8cm}
    \centering
    \includegraphics[width=0.9\textwidth]{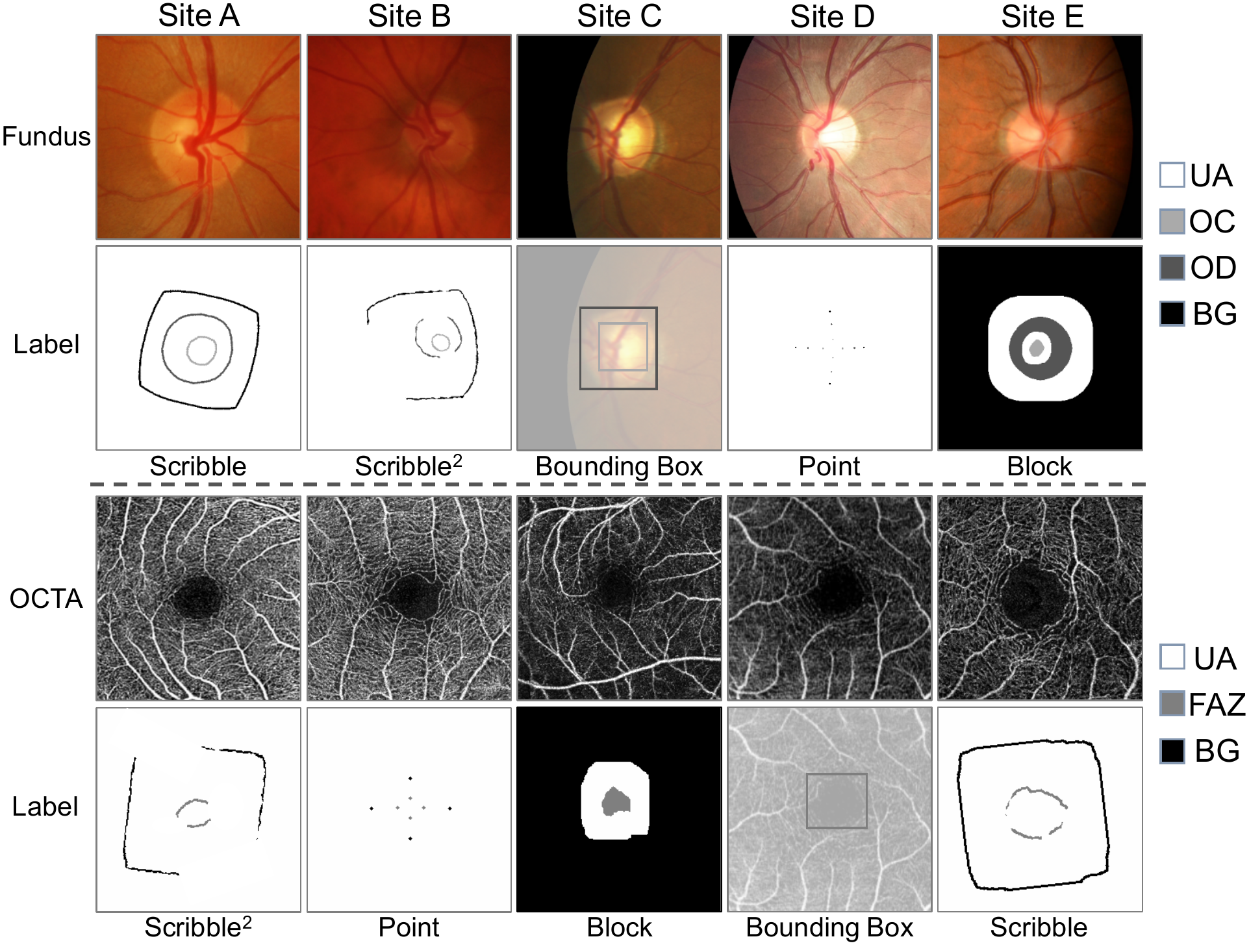}
    \caption{Examples of the image and the corresponding sparse annotation from each site. UA, OC, OD, FAZ, and BG respectively represent unlabeled area, optic cup, optic disc, foveal avascular zone, and background.} \label{figa1}
    \vspace{-0.6cm}
\end{figure}  
\begin{figure}[!h]
    \centering
    \includegraphics[width=0.4\textwidth]{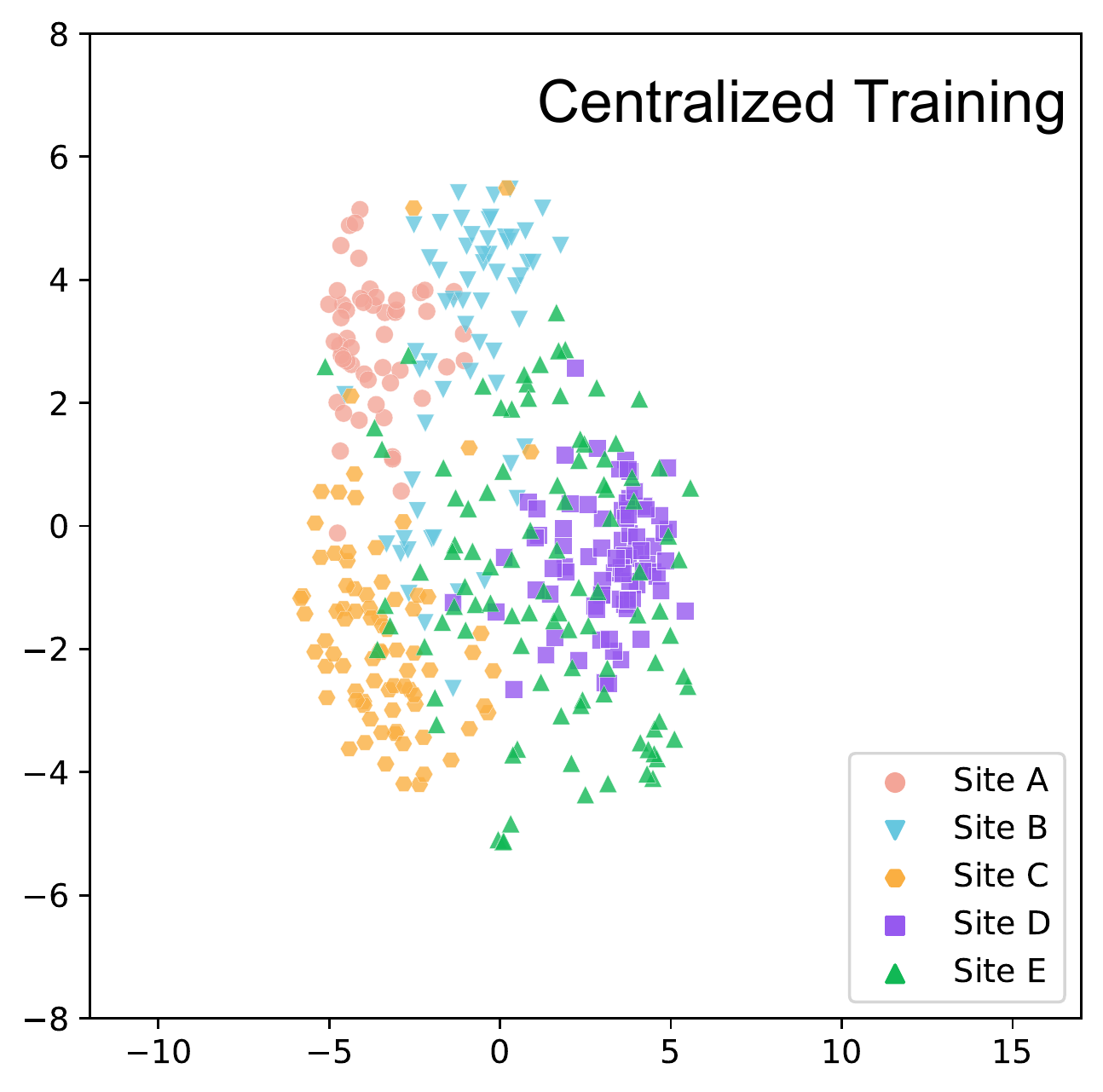}
    \hspace{+0.2cm}
    \includegraphics[width=0.4\textwidth]{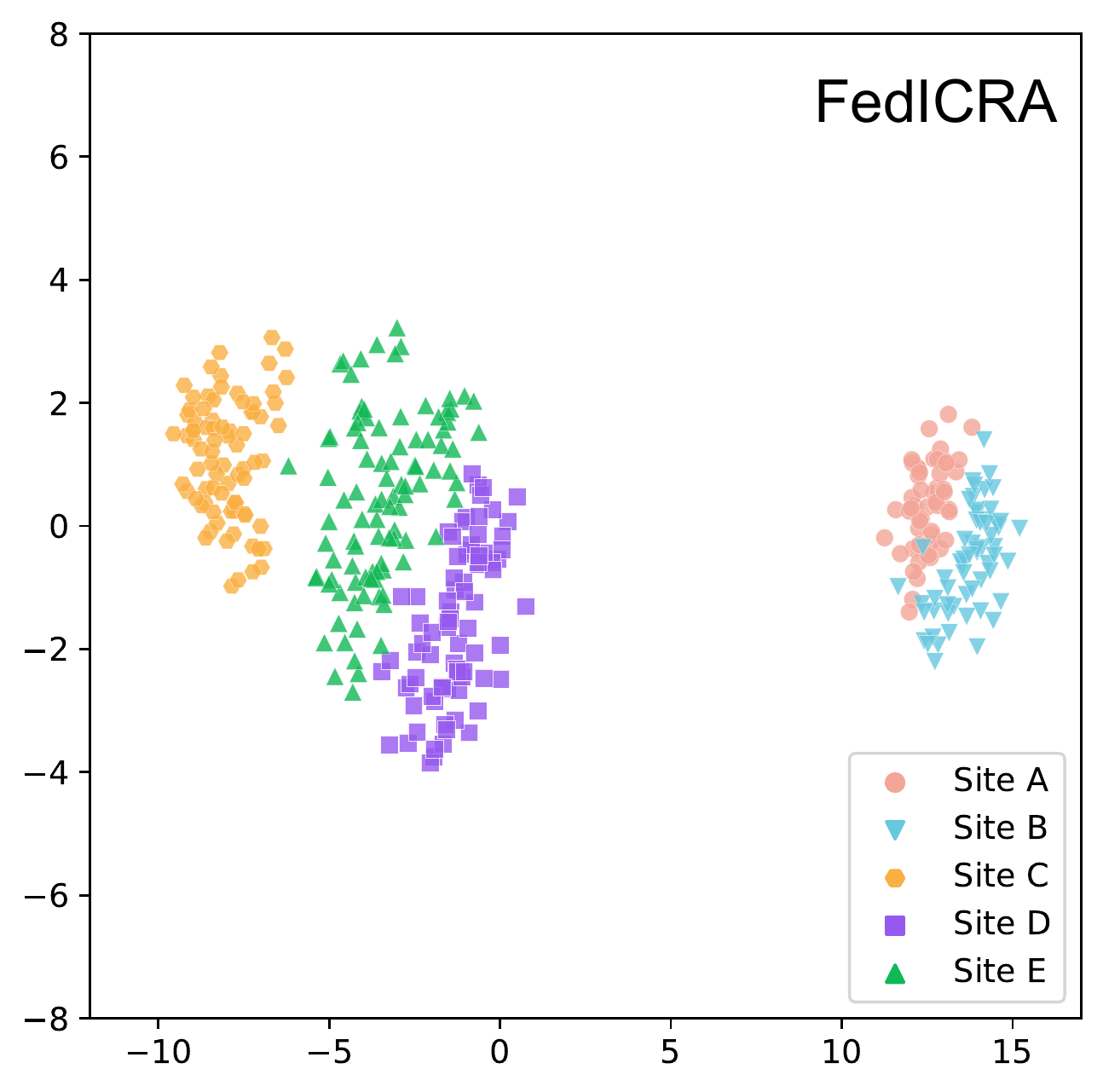}
    \vspace{-0.2cm}
    \caption{Comparison of data embeddings on ODOC segmentation test sets via t-SNE visualization.} \label{figa2}
    \vspace{-0.2cm}
\end{figure}

\begin{table}[!h]
    \vspace{-0.4cm}
    \caption{
        Performance comparisons (HD95) of different FL methods as well as different local and centralized training settings on ODOC segmentation. The best results are highlighted in bold and the second-best ones are underlined.}
    \label{tablea2}
    \setlength{\tabcolsep}{0.6mm}
    \vspace{+0.2cm}
    \centering
    \resizebox{\textwidth}{!}{
        \begin{tabular}{l|ccccc|c|ccccc|c|c}
            \specialrule{0.12em}{0pt}{0pt}
            & \multicolumn{6}{c|}{HD95 (OD) $\downarrow$}                                                    & \multicolumn{6}{c|}{HD95 (OC)$\downarrow$}                                                     & \multirow{2}{*}{Overall}        \\
            \cline{1-13}
Methods          & Site A          & Site B           & Site C          & Site D           & Site E           & Avg.  & Site A           & Site B           & Site C          & Site D           & Site E           & Avg.  &  \\
\specialrule{0.05em}{0pt}{0pt}
FedAvg         & 13.17±8.04  & 24.10±26.68 & 9.28±7.23    & 20.80±43.31  & 24.59±33.12 & 18.39 & 18.56±12.99 & 23.98±28.08 & \underline{11.19±9.99}  & 22.42±47.48 & 16.80±32.37 & 18.59 & 18.49 \\
FedProx        & \underline{8.38±4.17}   & \underline{15.72±21.67} & 10.64±14.23  & 16.02±27.82  & 20.36±31.30 & 14.22 & \textbf{14.17±6.91}  & 12.82±7.34  & 13.36±13.58 & 13.91±25.36 & 16.10±30.24 & 14.07 & 14.15 \\
FT             & 10.07±8.96  & 22.26±25.08 & \textbf{8.51±7.58}    & 40.12±53.19  & 20.36±31.19 & 20.26 & 19.96±12.44 & 22.24±27.52 & 11.68±9.94  & 31.60±51.63 & 14.59±26.15 & 20.02 & 20.14 \\
FedBN         & 8.78±4.77   & 18.06±28.02 & 19.54±32.51  & \underline{10.72±25.22}  & 17.68±28.85 & 14.96 & 15.70±7.14  & \underline{12.08±6.79}  & 17.24±32.24 & 9.96±24.07  & \underline{11.96±13.18} & 13.39 & 14.17 \\
FedAP           & 9.19±6.18   & 21.59±14.86 & \underline{8.96±12.14}   & \textbf{5.77±2.61}    & \underline{16.80±29.82} & \underline{12.46} & 15.69±7.47  & 14.68±19.84 & 12.66±14.05 & \underline{7.05±3.19}   & 14.49±25.30 & \underline{12.91} & \underline{12.69} \\
FedRep          & 17.39±28.48 & 21.59±27.13 & 23.56±34.86  & 54.10±68.55  & 31.00±39.85 & 29.53 & 27.61±37.21 & 18.90±21.65 & 17.89±29.64 & 50.60±79.80 & 17.71±28.70 & 26.54 & 28.03 \\
MetaFed       & 41.99±52.18 & 28.19±44.51 & 119.84±70.20 & 33.57±55.59  & 23.06±20.78 & 49.33 & 34.82±43.31 & 13.86±7.60  & 32.83±53.45 & 9.37±14.39  & 14.56±16.37 & 21.09 & 35.21 \\
FedICRA (Ours) & \textbf{7.83±5.20}   & \textbf{10.36±18.40} & 10.10±9.81    & 10.87±20.65    & \textbf{9.02±10.84}  & \textbf{9.64}  & \underline{15.47±10.27} & \textbf{11.86±6.37} & \textbf{8.61±3.88}   & \textbf{6.13±3.10}   & \textbf{8.47±9.29}  & \textbf{10.11} & \textbf{9.87} \\
\specialrule{0.05em}{0pt}{0pt}
LT (Weak)\cellcolor{gray!30}      & 26.91±42.21\cellcolor{gray!30} & 44.05±59.07\cellcolor{gray!30} & 74.35±74.73\cellcolor{gray!30}  & 151.71±54.14\cellcolor{gray!30} & 19.53±23.93\cellcolor{gray!30} & 63.31\cellcolor{gray!30} & 16.51±8.51\cellcolor{gray!30}  & 35.98±58.07\cellcolor{gray!30} & 24.93±43.10\cellcolor{gray!30} & 19.27±33.63\cellcolor{gray!30} & 16.45±22.63\cellcolor{gray!30} & 22.62\cellcolor{gray!30} & 42.97\cellcolor{gray!30} \\
CT (Weak)\cellcolor{gray!30}       & 9.32±5.86\cellcolor{gray!30}   & 12.25±7.00\cellcolor{gray!30}  & 7.29±3.23\cellcolor{gray!30}    & 11.97±15.68\cellcolor{gray!30}  & 14.52±23.10\cellcolor{gray!30} & 11.07\cellcolor{gray!30} & 15.55±7.30\cellcolor{gray!30}  & 13.79±7.13\cellcolor{gray!30}  & 10.73±4.32\cellcolor{gray!30}  & 9.83±16.04\cellcolor{gray!30}  & 12.15±19.02\cellcolor{gray!30} & 12.41\cellcolor{gray!30} & 11.74\cellcolor{gray!30} \\
LT (Full)\cellcolor{gray!30}      & 9.27±10.42\cellcolor{gray!30}  & 6.26±4.83\cellcolor{gray!30}   & 7.55±6.05\cellcolor{gray!30}    & 5.38±2.63\cellcolor{gray!30}    & 11.07±20.82\cellcolor{gray!30} & 7.91\cellcolor{gray!30}  & 13.34±7.13\cellcolor{gray!30}  & 12.62±8.56\cellcolor{gray!30}  & 9.18±5.43\cellcolor{gray!30}   & 6.30±3.48\cellcolor{gray!30}   & 11.51±16.56\cellcolor{gray!30} & 10.59\cellcolor{gray!30} & 9.25\cellcolor{gray!30}  \\
CT (Full)\cellcolor{gray!30}      & 6.40±3.48\cellcolor{gray!30}   & 8.95±7.34\cellcolor{gray!30}   & 6.13±2.53\cellcolor{gray!30}    & 4.62±2.01\cellcolor{gray!30}    & 8.92±18.19\cellcolor{gray!30}  & 7.00\cellcolor{gray!30}     & 12.31±5.45\cellcolor{gray!30}  & 11.28±6.95\cellcolor{gray!30}  & 8.47±4.47\cellcolor{gray!30}   & 5.74±2.70\cellcolor{gray!30}   & 8.48±11.91\cellcolor{gray!30}  & 9.26\cellcolor{gray!30}  & 8.13\cellcolor{gray!30}  \\
\specialrule{0.12em}{0pt}{0pt}

\end{tabular}
    }
\end{table}

\begin{figure}[H]
    \vspace{-0.4cm}
    \includegraphics[width=\textwidth]{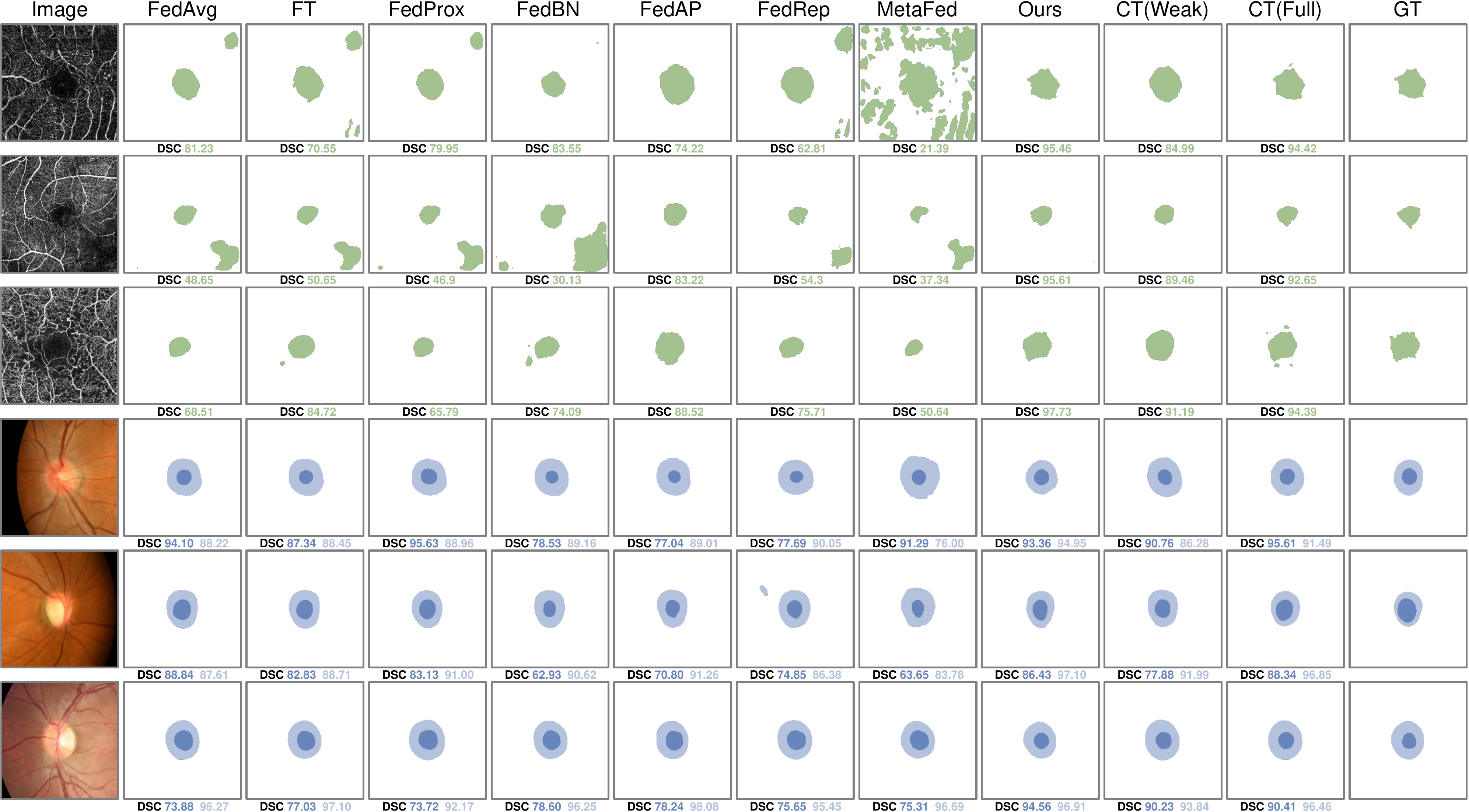}
    \vspace{-0.6cm}
    \caption{Visualization results from FedICRA and other SOTA methods (\textbf{CT} indicates centralized training, with \textbf{Weak} and \textbf{Full} respectively denoting utilizing sparse annotations and full masks).} \label{figa3}
\end{figure}

\end{document}